\documentclass{iopart}
\usepackage{graphicx}

\begin{document}

\letter{Strongly correlated wave functions for artificial atoms and molecules}

\author{Constantine Yannouleas and Uzi Landman} 
\address{
School of Physics, Georgia Institute of Technology,
Atlanta, Georgia 30332-0430 }


\begin{abstract}
A method for constructing semianalytical strongly correlated wave functions 
for single and molecular quantum dots is presented. It employs a two-step 
approach of symmetry breaking at the Hartree-Fock level and of subsequent 
restoration of total spin and angular momentum symmetries via Projection 
Techniques. Illustrative applications are presented for the case of a 
two-electron helium-like single quantum dot and a hydrogen-like quantum dot 
molecule.\\
~~~~~~\\
~~~~~~~\\
Published: J. Phys.: Condens. Matter {\bf 14}, L591 (2002)
\end{abstract}


\maketitle

Understanding the nature of strong many-body correlations in condensed-matter
finite systems is a fundamental task, which can be facilitated by devising 
analytical or semi-analytical many-body wave functions that approximate
well the exact solutions and capture the essential physical 
properties of such systems. The description of strong correlations 
in the two-dimensional (2D) electron gas in a high magnetic field by the 
Laughlin wave function \cite{lau} represents a celebrated example of such a 
methodology.

Here we construct semianalytical correlated wave functions for electrons in
2D single and molecular Quantum Dots (QD's), which are manmade devices created
at semiconductor interfaces and are most often referred to as artificial 
atoms and molecules (since they contain a finite number of electrons).
To this end, we employ a two-step approach 
consisting of symmetry breaking at the Hartree-Fock single-determinantal level
and subsequent symmetry restoration via Projection Techniques (PT's) resulting
in multi-determinantal wave functions. PT's have been introduced earlier in 
Quantum Chemistry \cite{low} for the restoration of the total spin
of a molecule and in Nuclear Physics \cite{py,sr} for the restoration of the 
total 3D angular momentum (space rotational symmetry) of deformed open-shell 
nuclei. Our application of these methods to circular single quantum dots
(SQD's) requires the simultaneous restoration of both the spin and the 
angular-momentum symmetries. This requirement poses a more demanding
challenge compared to the task of restoring a single broken symmetry,
as is the case with the Quantum-Chemistry and Nuclear-Physics many-body 
problems.
 
Since the approach of restoration of broken symmetries is largely untested 
and unknown in the framework of 2D  QD's, we aim in this paper at focusing on
methodological aspects. For simplicity and conceptual clarity, we construct 
below strongly correlated wave functions for the case of two interacting 
electrons in a SQD (artificial helium, He-QD) and in a lateral double 
quantum dot (artificial hydrogen quantum dot molecule, H$_2$-QDM). 
Two strongly-correlated electrons (with or without an external magnetic field)
may exhibit a complex behavior \cite{lau,kh,yl1} and can provide the 
foundation for understanding the properties of a larger number of interacting 
electrons. Furthermore, the recent growth of interest in quantum computing
focussed attention on the potential ability to manipulate the ground-state
entanglement of two-electron quantum-dot systems \cite{burk}.

The two-body hamiltonian for two interacting electrons constrained to move
on a plane is given by,
\begin{equation}
{\cal H} = H({\bf r}_1)+H({\bf r}_2)+e^2/\kappa r_{12}~,
\label{ham}
\end{equation}
where the last term is the Coulomb repulsion with $\kappa$ being the
dielectric constant of the semiconductor material. $H({\bf r})$ is the 
single-particle hamiltonian for an electron in an external perpendicular 
magnetic field ${\bf B}$ and an appropriate potential confinement. 
For a QDM, the external confinement is given by a two-center-oscillator 
potential, and the single-particle hamiltonian is written as,
\begin{equation}
H=T + \frac{1}{2} m^* \omega^2_{0} (x^2 + y^{\prime 2}_k)
    + V_{\rm {neck}}(y) + \frac{g^* \mu_B}{\hbar} {\bf B \cdot s}~,
\label{hsp}
\end{equation} 
where $y_k^\prime=y-y_k$ with $k=1$ for $y<0$ (left) and $k=2$ for $y>0$ 
(right). $x$ denotes the coordinate perpendicular to the interdot axis
($y$). $T=({\bf p}-e{\bf A}/c)^2/2m^*$, with ${\bf A}=0.5(-By,Bx,0)$, and the 
last term in Eq. (\ref{hsp}) is the Zeeman interaction with $g^*$ being the 
effective $g$ factor, $\mu_B$ the Bohr magneton, and ${\bf s}$ the spin
of an individual electron. The shapes described by $H$ are two equal
semi-circles connected by a smooth neck [$V_{\rm{neck}}(y)$]. $-y_1 = y_2 >0$  
are the centers of these semi-circles, $d=y_2-y_1$ is the interdot 
distance, and $m^*$ is the effective electron mass. The case of a SQD is 
obtained for $d=0$, and in this case the confining potential is purely 
parabolic. For the smooth neck, we use  
$V_{\rm{neck}}(y) = \frac{1}{2} m^* \omega^2_{0} 
[C_k y^{\prime 3}_k + D_k y^{\prime 4}_k] \theta(|y|-|y_k|)$, 
where $\theta(u)=0$ for $u>0$ and $\theta(u)=1$ for $u<0$.
The constants $C_k$ and $D_k$ can be expressed via one parameter,
as follows: $C_k = (2-4\epsilon^b)/y_k$ and $D_k=(1-3\epsilon^b)/y_k^2$, 
where the barrier-control parameter $\epsilon^b=V_b/V_{0}$ 
is related to the actual (controlable) height 
of the bare barrier ($V_b$) between the two QD's, and 
$V_{0}=m^* \omega_{0}^2 y_1^2/2$. 
The single-particle levels of $H$ are obtained by numerical diagonalization 
(for details see Ref. \cite{yl4}).

In each case, in the first step of our procedure, the two-electron hamiltonian 
[Eq.\ (\ref{ham})] is solved \cite{yl2,yl3,yl4} in the (symmetry-breaking) 
spin-and-space unrestricted Hartree-Fock (sS-UHF) approximation; for 
comparison, the symmetry-adapted restricted Hartree-Fock (RHF) will also be 
considered. In all cases, we will use $\hbar \omega_0 =5$ meV and
$m^*=0.067 m_e$ (this effective-mass value corresponds to GaAs).
We will vary the dielectric constant $\kappa$, however, in order to control
the ratio of the strength of the Coulomb repulsion relative to the zero-point 
kinetic energy (see in particular the case of the SQD and figure 3 below).
The variation of this ratio provides us with the ability to study the whole 
range of electron correlations, from the regime of weak correlations to that of
strong correlations. 

We start with the case of a lateral H$_2$-QDM in a magnetic field, where only 
the spin projection needs to be considered, since the confining potential 
lacks circular symmetry. The example we discuss here\footnote{
For a systematic study of the H$_2$-QDM at different values of $\kappa$
[including $\kappa=12.9$ (GaAs)] and interdot barrier heights,
demonstrating gate control of the entanglement of a pair of electrons, see
Ref.\ \cite{yl4}.} 
~corresponds to the case of weak interelectron repulsion 
(the full choice of parameters is given in
the caption of Fig.\ 1). The sS-UHF determinant (henceforth we drop the prefix
sS in subscripts) which describes the ``singlet'' (see below) ground state of 
the H$_2$-QDM is given by the expression,
\begin{equation}
|\Phi_{\rm{UHF}}(1,2)\rangle = 
| u(1) \bar{v}(2) \rangle /\sqrt{2}~.
\label{det}
\end{equation}
In Eq.\ (\ref{det}), we have used a compact notation for the Slater 
determinant with $u(1) \equiv u({\bf r}_1) \alpha(1)$ and
$\bar{v}(2) \equiv v({\bf r}_2) \beta(2)$, where $u({\bf r})$ and $v({\bf r})$ 
are the $1s$-like (left) and $1s^\prime$-like (right) localized orbitals of 
the sS-UHF solution, and $\alpha$ and $\beta$ denote the up and down spin 
functions, respectively. Such orbitals for the field-free case are displayed 
in Fig.\ 1(a) (left column). Similar localized orbitals (which are complex 
functions) can appear also in the $B \neq 0$ case [see Fig.\ 1(b), right 
column].

$|\Phi_{\rm{UHF}}(1,2)\rangle$ is an eigenstate of the $z$-projection of the 
total spin, $\hat{{\bf S}} = \hat{{\bf s}}_1 + \hat{{\bf s}}_2$, with 
eigenvalue $S_z=0$. However, it is not an eigenstate of the square, 
$\hat{{\bf S}}^2$, 
of the total spin (thus the quotation marks in ``singlet'' 
above). From the determinant $|\Phi_{\rm{UHF}}(1,2)\rangle$, one can generate 
a singlet eigenstate of $\hat{{\bf S}}^2$ (with zero eigenvalue) by applying 
the projection operator $P_0 \equiv 1 - \varpi_{12}$,
where the operator $\varpi_{12}$ interchanges the spins of the two electrons.

Thus the singlet state of the two localized electrons is given by the 
projected wave function, 
\begin{equation}
|\Psi(1,2)\rangle \equiv P_0 |\Phi_{\rm{UHF}}(1,2)\rangle \propto  
| u(1) \bar{v}(2) \rangle - \;| \bar{u}(1) v(2) \rangle~.
\label{prj0}
\end{equation}
In contrast to the single-determinantal wave functions of the RHF and sS-UHF 
methods, the projected many-body wave 
function (\ref{prj0}) is a linear superposition of two Slater determinants, 
and thus it is an entangled state representing a corrective 
(post-Hartree-Fock) step beyond the mean-field approximation. We notice that 
the spatial reflection symmetry is automatically restored along with the spin 
symmetry.

Eq.\ (\ref{prj0}) has the form of a Heitler-London (HL) or valence bond 
\cite{hel} wave function. However, unlike the original HL scheme which uses 
the orbitals $\phi_L({\bf r})$ and $\phi_R({\bf r})$ of the separated 
(left and right) QD's, expression (\ref{prj0}) employs
the sS-UHF orbitals which are self-consistently optimized for any separation 
$d$, ineterdot barrier height $V_b$, and magnetic field $B$.
Consequently, expression (\ref{prj0}) can be characterized as a Generalized 
Heitler-London (GHL) wave function. In the context of QD's,
the simple HL approach has been proven very useful in demonstrating 
\cite{burk} that the H$_2$-QDM under an external magnetic field can function 
as an elemental two-qubit logic gate. Our more accurate GHL approach has the 
potential of greatly improving the quality of similar investigations 
regarding the implementation of quantum computing using solid-state 
nanodevices. 

We further notice that our GHL method belongs to a class of projection
techniques known as variation before projection (VBP), unlike the familiar in
Quantum Chemistry Generalized Valence Bond method \cite{god}, or the 
Spin-Coupled Valence Bond method \cite{gerr}, which employ a variation after 
projection (VAP) (see e.g. ch. 11.4.2 of Ref.\ \cite{sr}). In the context of 
QD's, a generalization of our GHL approach along the VAP technique may provide
even more accurate results. This, however, is left for future work, including 
the development of the pertinent computer codes.

The energy of a projected state is given \cite{sr} in general by the formula,
\begin{equation}
E_{\rm{PROJ}} = \left. 
\langle \Phi_{\rm{UHF}}|{\cal H} {\cal O} |\Phi_{\rm{UHF}}\rangle
\right/ \langle \Phi_{\rm{UHF}}|{\cal O}|\Phi_{\rm{UHF}}\rangle~,
\label{epr}
\end{equation}
where ${\cal H}$ is the many-body hamiltonian and ${\cal O}$ is a general 
projection operator with the property ${\cal O}^2={\cal O}$. 

Using the spin-projection operator $P_0$ in place of ${\cal O}$, we find for 
the total energy, $E^{\rm{s}}_{\rm{GHL}}$, of the singlet GHL state,
\begin{equation}
E^{\rm{s}}_{\rm{GHL}}={\cal N}^2_{\rm{s}}
[ H_{uu}+H_{vv} +
 S_{uv}H_{vu} + S_{vu}H_{uv}+J_{uv} + K_{uv}]~,
\label{engvb}
\end{equation}
where $H_{uu}$, etc., are the matrix elements of the single-particle 
hamiltonian $H$ in Eq.\ (\ref{ham}), and $J$ and $K$ are the direct and 
exchange matrix elements of $e^2/\kappa r_{12}$. 
$S_{uv}$ is the overlap integral of the $u({\bf r})$ and $v({\bf r})$ 
orbitals, 
\begin{equation}
S_{uv}= \int u^*({\bf r})v({\bf r}) d{\bf r}~,
\label{suv}
\end{equation}
and the normalization parameter is given by
\begin{equation}
{\cal N}^2_{\rm{s}} = 1/(1 + S_{uv}S_{vu})~.
\label{nuv}
\end{equation}

At zero magnetic field, the electron orbitals are real functions and
Eqs.\ (\ref{engvb}) $-$ (\ref{nuv}) reduce to a form familiar from the
Generalized Valence Bond theory of Quantum Chemistry \cite{god}.

For the triplet state with $S_z=\pm 1$, the projected wave function coincides 
with the original HF determinant, so that the corresponding energies in all
three approximation levels are equal, i.e., 
$E^{\rm{t}}_{\rm{GHL}}=E^{\rm{t}}_{\rm{RHF}}=
E^{\rm{t}}_{\rm{UHF}}$.
 
\begin{figure}[t]
\vspace*{3.cm}
\begin{center}
{\bf FIGURE 1}\\
{\bf SEPARATE GIF}
\end{center}
\vspace*{3.cm}
~~~~~~~~~~~\\
\caption{
H$_2$-QDM: The two occupied orbitals (modulus square) 
of the symmetry broken ``singlet''
sS-UHF solution for (a) $B=0$ and (b) $B=9$ T. (c) The singlet-triplet 
energy difference as a function of an external perpendicular magnetic field 
and for three successive levels of approximation, i.e., the RHF (top solid 
curve), the sS-UHF (dashed curve), and the GHL (bottom solid curve). 
The choice of parameters is: parabolic confinement of each dot $\hbar \omega_0=
5$ meV, interdot distance $d=30$ nm, interdot barrier height $V_b=3.71$ meV,
effective mass $m^*=0.067 m_e$, and dielectric constant $\kappa=45$. The 
effective Zeeman coefficient was chosen $g^*=0$; for a small nonvanishing 
value, like $g^*=-0.44$, the Zeemann splitting does not alter the orbital 
densities and can be simply added to $\Delta \varepsilon$. Up and down arrows 
denote spins. Distances are in nm and the densities in 10$^{-4}$ nm$^{-2}$.
}
\end{figure}

In Fig.\ 1(c), we display the singlet-triplet energy gap,
$\Delta \varepsilon = E^{\rm{s}}-E^{\rm{t}}$, 
of the H$_2$-QDM as a function of the magnetic field $B$.
For all three levels of approximation,
$\Delta \varepsilon$ starts from a negative minimum (singlet ground state) and
after crossing the zero value it remains positive (triplet ground state).
However, after crossing the zero line, the RHF curve incorrectly continues
to rise sharply and moves outside of the plotted range. After reaching a
broad maximum, the positive $\Delta \varepsilon$ branches of both the
sS-UHF and GHL curves approach zero for sufficiently large $B$,
a behavior which indicates that the H$_2$-QDM dissociates with $B$.
This molecular dissociation can be further seen in the orbitals themselves: 
at $B=9$ T [Fig.\ 1(b)] the orbitals are well localized on the individual
dots, while at $B=0$ [Fig.\ 1(a)] they extend over the entire QDM.
Note that, in addition to having the proper symmetry, GHL is energetically
the best approximation.

We further note that systematic explorations for determining the border
of HF instability (i.e., the appearance of broken symmetry solutions with
lower energy) in the case of the H$_2$-QDM are not available. At $B=0$,
this border depends on all four parameters $\hbar \omega_0$, $\kappa$, $d$, 
and $V_b$. For $\hbar \omega_0 = 5$ meV, $d=30$ nm, and $\kappa = 45$, it was 
found \cite{yl4} that this border can be crossed by reducing the interdot
barrier $V_b$ to zero. For the same values of $B$, $\hbar \omega_0$ and
$d$, but for a different $\kappa =12.9$ (stronger interelectron repulsion,
as is the case for GaAs), however, all values of the interdot barrier ($V_b 
\geq 0$) represent cases that lie well within the region of instability.

Next we address the case of two electrons at B=0 in a single QD with a
parabolic confinement. As we have shown in earlier publications 
\cite{yl2,yl3}, the sS-UHF solution does not preserve the rotational symmetry 
for $R_W \geq 1.0$ (for $R_W < 1.0$, the sS-UHF solution collapses onto the
RHF one). The Wigner parameter $R_W$ expresses the ratio between
the interelectron repulsion $Q$ and the zero-point kinetic energy $K$;
it is customary to take $Q=e^2/\kappa l_0$ and $K \equiv \hbar \omega_0$,
where $l_0=(\hbar/m^* \omega_0)^{1/2}$ is the spatial extent of an electron 
in the lowest state of the parabolic confinement. This gives $R_W \propto
1/(\kappa \sqrt{\omega_0})$, showing that it can be varied through the
choice of materials (i.e., $\kappa$) and/or the strength of the confinement
($\omega_0$). For $R_W \geq 1.0$ and for $S_z=0$, the sS-UHF yields 
\cite{yl3} an infinite manifold of azimuthally degenerate ground-state 
configurations consisting of two localized antipodal orbitals, suggesting 
formation of an electron molecule \cite{yl1,yl2,yl3} (also referred to as
Wigner molecule). For $\kappa=8$ ($R_W=2.39$) 
and for a specific azimuthal direction (i.e., $\gamma=0$), the two localized 
orbitals, $u ({\bf r})$ and $v ({\bf r})$ with up and down spin, respectively,
are depicted in Fig.\ 2(a).
As was the case with the double-dot example studied above, the $S_z=0$ sS-UHF 
determinant $|\Phi_{\rm{UHF}}(1,2) \rangle$ for $\gamma=0$ is given by 
Eq.\ (\ref{det}) and thus, in addition to the rotational symmetry, it does 
not preserve the total-spin symmetry. For the singlet state, one can 
restore both symmetries successively, namely, one can generate appropriate 
projected wave functions,
\numparts
\begin{equation}
|\Psi_I (1,2)\rangle \equiv {\cal O} |\Phi_{\rm{UHF}}(1,2)\rangle~,
\label{psisam}
\end{equation}
by applying the product operator,\footnote{
In the case of a triplet state, the sS-UHF determinant with $S_z= \pm 1$
preserves the total spin, and thus the application of ${\cal P}_I$ alone is
sufficient.}
\begin{equation}
{\cal O} \equiv {\cal P}_I P_0~,
\label{cpr}
\end{equation}
\label{both}
\endnumparts
\noindent
where the spin-projection operator $P_0$ produces a two-determinant singlet 
GHL wave function as previously explained. The angular-momentum projection
operator ${\cal P}_I$ yields multideterminantal wave functions having good 
total angular momentum $I$. This latter operator produces a linear 
superposition of an infinite number of azimuthal 
GHL wave functions and it is given\footnote{
The corresponding expression \cite{py} for the 3D angular-momentum projection
uses the Wigner functions ${\cal D}^I_{MK}(\Omega)$ (see also ch. 11.4.6 in 
Ref.\ \cite{sr}).}
~by \cite{sr},
\begin{equation}
2 \pi {\cal P}_I \equiv \int_0^{2 \pi} 
d\gamma \exp[-i \gamma (\hat{L}-I)]~,
\label{amp}
\end{equation} 
where $\hat{L}=\hat{l}_1+\hat{l}_2$ is the operator for the total angular 
momentum. In the following we focus on the ground state\footnote{
The family of projected wave functions (9) describes all the
lowest-energy (yrast band \cite{yl1}) states with good angular momentum
$I=2, 4, ...$, in addition to the ground state ($I=0$).
The yrast-band states with odd values, $I=$ 1, 3, 5, ..., are generated
via a projection of the triplet state.}
~of the system with $I=0$.

\begin{figure}[t]
~~~~~~~~~~~\\
\vspace*{3.cm}
\begin{center}
{\bf FIGURE 2}\\
{\bf SEPARATE GIF}
\end{center}
\vspace*{3.cm}
~~~~~\\
\caption{
He-QD at $B=0$: (a) The two occupied orbitals (modulus aquare) of the 
symmetry-broken ``singlet'' sS-UHF solution. The rest of the frames display 
the electron densities at successive approximation levels, i.e., (b) RHF,
(c) sS-UHF, (d) SP alone, (e) S\&AMP, and (f) exact. See the text for 
a description of the acronyms. The choice of parameters is:
dielectric constant $\kappa=8$, parabolic confinemenet $\hbar \omega_0=5$ meV, 
and effective mass $m^*=0.067 m_e$.
Distances are in nm and the densities in 10$^{-4}$ nm$^{-2}$.
}
\end{figure}

It is instructive to examine the transformations of the ground-state electron 
densities (ED's) resulting from the successive approximations,
RHF, sS-UHF, spin projection (SP), and combined spin and angular momentum 
projection (S\&AMP). For $\kappa=8$, such ED's are shown in Fig.\ 2. 
Since the exact solution for two electrons in a 
parabolic confinement is available \cite{yl1}, we also plot the
corresponding ED for the exact ground state in Fig.\ 2(f). 
The ED's of the initial RHF [Fig.\ 2(b)] and the final S\&AMP [Fig.\ 2(e)] 
approximations are circularly symmetric, while those of the two 
intermediate approximations, i.e., the sS-UHF and SP, do break the circular 
symmetry. This behavior graphically illustrates the meaning of the term
restoration of symmetry and the interpretation that the sS-UHF broken-symmetry
solution refers to the {\it intrinsic\/} (rotating) frame of reference of
the electron molecule. We notice that the S\&AMP electron density 
exhibits a characteristic flattening at the top in contrast to the more 
Gaussian-type RHF one. Thus, although not identical (the exact ED is slightly
flatter at the top), the S\&AMP ED closely resembles the exact one displayed
in Fig.\ 2(f). Further, we remark that the SP electron density exhibits in the
middle a shallower depression than the corresponding sS-UHF one, in keeping 
with the fact that the covalent bonding increases the probability of finding 
the electrons between the individual dots.
 
Using the projection operator (\ref{cpr}) in the general Eq.\ (\ref{epr}),
we obtain for the energy $E_{\rm{S\&AMP}}$ of the fully projected ground 
state of the two-electron single QD,
\begin{equation}
E_{\rm{S\&AMP}} = \left. { \int_0^{2\pi} h(\gamma) d\gamma } \right/
{ \int_0^{2\pi} n(\gamma) d\gamma}~,
\label{eproj}
\end{equation}
with
\begin{equation}
h(\gamma) = 
H_{us}S_{vt}+ H_{ut}S_{vs}+ H_{vt}S_{us}+H_{vs}S_{ut}+ 
 V_{uvst} + V_{uvts}~, 
\label{hgam}
\end{equation}
and
\begin{equation}
n(\gamma)= S_{us}S_{vt}+S_{ut}S_{vs}~,
\label{ngam}
\end{equation}
$s({\bf r})$ and $t({\bf r})$ being the $u({\bf r})$ and $v({\bf r})$
sS-UHF orbitals rotated by an angle $\gamma$, respectively. 
$V_{uvst}$ and $V_{uvts}$ are two-body matrix elements of the Coulomb 
repulsion.\footnote{
$V_{uvst} \equiv (e^2/\kappa) \int d{\bf r}_1 \int d{\bf r}_2
u^*({\bf r}_1) v^*({\bf r}_2) (1/r_{12}) s({\bf r}_1) t({\bf r}_2)$.
In Eq. (\ref{engvb}) $J_{uv} = V_{uvuv}$ and $K_{uv} = V{uvvu}$.}
~Observe that $E_{\rm{S\&AMP}}$ represents a rotational
average over all the possible orientations of the electron molecule and that 
$h(0)/n(0)$ coincides with the expression (\ref{engvb}) for the energy
$E^{\rm{s}}_{\rm{GHL}}$ of the covalently bonded H$_2$-QDM.

To further examine how closely expression (9) describes the
ground state of the He-QD, we display in Fig.\ 3 the energy deviations
of the four approximation steps (i.e., RHF, sS-UHF, SP, and S\&AMP) from the 
exact (EX) \cite{yl1} ground state, as a function of the dielectric
constant $\kappa$. The variation of $\kappa$ produces a 
variation in the strength of the interelectron repulsion,
with a larger repulsion (i.e., smaller $\kappa$) corresponding to stronger
electron correlations. 

\begin{figure}[t]
\centering\includegraphics[width=6.7cm]{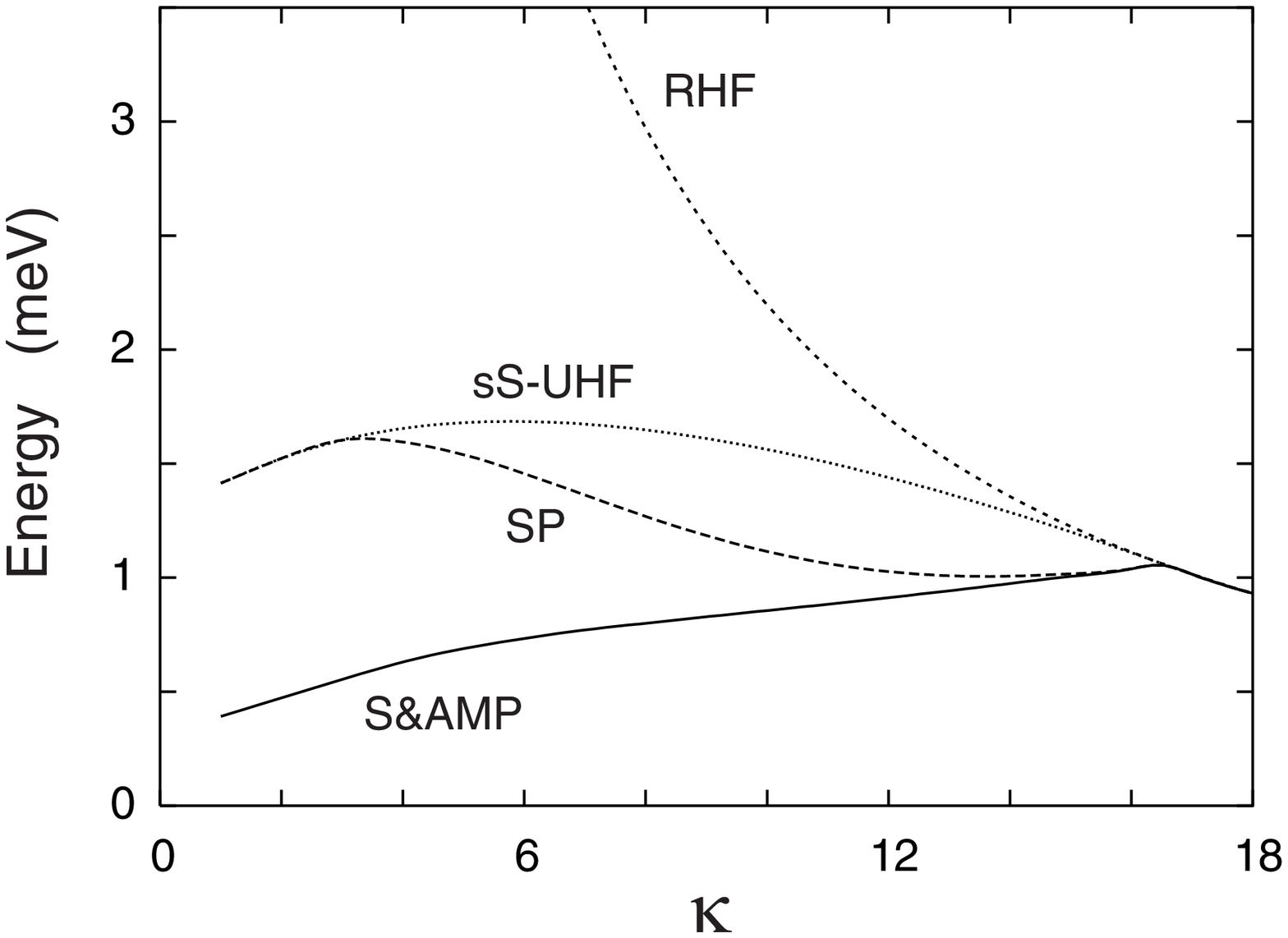}
~~~~~~~~~~~\\
\caption{
He-QD at $B=0$: Difference between the ground-state energies of various 
approximations and the exact one, plotted vs. the dielectric constant
$\kappa$. From top to bottom, the curves correspond to the RHF, the sS-UHF, 
the SP alone, and the combined S\&AMP approximation. The curve labeled RHF
gives the correlation energy. The choice of parameters
is: parabolic confinement $\hbar \omega_0=5$ meV, and effective mass 
$m^*=0.067 m_e$.
}
\end{figure}

The correlation energy, 
$E_{\rm{cor}} \equiv E_{\rm{RHF}} - E_{\rm{EX}}$, 
is defined as the difference between the RHF ground-state energy and the exact
one, and thus it coincides with the top curve in Fig.\ 3 denoted as RHF. 
The ordering (from top to bottom) of the curves in Fig.\ 3 reflects the fact 
that each subsequent approximation step captures successively a larger portion 
of the total correlations, and thus lowers the total energy. From Fig.\ 3, one 
sees that there are two correlation regimes: In the regime of weak 
correlations, the three lower curves collapse onto the RHF one; 
naturally, this regime corresponds to the normal Fermi liquid and
extends to $\kappa \rightarrow \infty$.
The onset of strong correlations and of symmetry-broken HF solutions occurs
at $\kappa = 16.5$ $(R_W = 1.16)$ and extends to $\kappa=0$.

Note that the percentage, $E_{\rm cor}/E_{\rm EX}$, of correlations
with respect to the exact energy is 6.8\% and 49.5\% for $\kappa=16.50$
and $\kappa=1$, respectively. These values are much higher than the
values encountered in natural atoms; for this reason, 2D QD's are referred to
as strongly correlated systems. Thus it is not surprising that the 
RHF error in Fig.\ 3 grows exponentially for stronger correlations (smaller
$\kappa$ or larger $R_W$). However, it is remarkable that the S\&AMP 
approximation (where all symmetries have been restored) converges to the
exact result for smaller $\kappa$. Note that the exact ground-state energy is 
19.80 meV for $\kappa=8$ ($R_W=2.39$) and 51.83 meV for $\kappa=1$ 
($R_W=19.09$), resulting for the S\&AMP in a relative error of 4\% and 0.7\%,
respectively.\footnote{
For $\kappa=8$, the fraction of correlation energy captured
by the successive approximations is: sS-UHF 44.4\%, SP 57.2\%, and S\&AMP
73.1\%; the corresponding values for $\kappa=1$ are:
94.5\%, 94.5\%, and 98.5\%.}
~These results are in agreement with the experience from Nuclear Physics,
where it has been found \cite{sr} that the VBP yields reliable results
in the case of strong symmetry breaking (strongly deformed nuclei).

Interestingly, the SP curve collapses onto the sS-UHF 
one for $\kappa \leq 3.2$. This behavior 
defines \cite{yl2} an intermediate regime\footnote{
In Ref.\ \cite{yl2}, we referred to the intermediate regime as the
weak-Wigner-molecule regime. Such an intermediate regime may also arise in the
infinite 2D electron gas at $B=0$, see Pichard J-L et al
{\it cond-mat/0107380}.}
~between the Fermi-liquid and 
the regime of strongly crystallized Wigner molecules ($\kappa \leq 3.2$).
In the latter regime, the overlap ($S_{uv}$) between the antipodally localized 
electron orbitals is very small, such that the sS-UHF energy is not 
effectively lowered by the spin projection (SP).

We have demonstrated that the sS-UHF method in conjunction
with the companion step of restoration of the spin and space
symmetries via Projection Techniques (when such symmetries are broken)
can provide appropriate semianalytical wave functions for an accurate
description of strongly correlated electrons in artificial atoms and molecules.

This research is supported by the US D.O.E. (Grant No. FG05-86ER-45234).


\section*{References}
\begin{thebibliography}{55}
\bibitem{lau}
Laughlin R B 
1983 {\it Phys. Rev. Lett.\/} {\bf 50} 1395;
1983 {\it Phys. Rev.\/} B {\bf 27} 3383
\bibitem{low}
L\"{o}wdin P O
1955 {\it Phys. Rev.\/} B {\bf 97} 1509; 
1964 {\it Rev. Mod. Phys.\/} {\bf 36} 966
\bibitem{py}
Peierls R E and Yoccoz J 
1957 {\it Proc. Phys. Soc., London, Sect. A\/}  {\bf 70} 381
\bibitem{sr}
Ring P and Schuck P
{\it The Nuclear Many-body Problem\/} (Spinger, New York, 1980)
ch. 11.4.4. 
\bibitem{kh}
Merkt U et al
1991 {\it Phys. Rev.\/} B {\bf 43} 7320
\bibitem{yl1}
Yannouleas C and Landman U
2000 {\it Phys. Rev. Lett.\/} {\bf 85} 1726
\bibitem{burk}
Burkard G et al
1999 {\it Phys. Rev.\/} B {\bf 59} 2070 
\bibitem{yl2}
Yannouleas C and Landman U
1999 {\it Phys. Rev. Lett.\/} {\bf 82} 5325; 
2000 {\it Ibid.\/} {\bf 85} (E)2220
\bibitem{yl3}
Yannouleas C and Landman U
2000 {\it Phys. Rev.\/} B {\bf 61} 15895
\bibitem{yl4}
Yannouleas C and Landman U 2001 {\it ArXiv: cond-mat/0109167}
\bibitem{hel}
Heitler H and London F
1927 {\it Z. Phys.\/} {\bf 44} 455
\bibitem{god}
Goddard III W A, Dunning, Jr. T H, Hunt W J and Hay P J
1973 {\it Acc. Chem. Res.} {\bf 6} 368
\bibitem{gerr}
Gerratt J and Lipscomb W N
1968 {\it Proc. Natl. Acad. Sci. (USA)\/} {\bf 59} 332 

\end{thebibliography}
\end{document}